\documentclass[11pt]{article}
\usepackage{moriond,epsfig}

\def\be{\begin{equation}}
\def\ee{\end{equation}}
\def\bea{\begin{eqnarray}}
\def\eea{\end{eqnarray}}

\usepackage{amsmath}

\usepackage{xspace}
\newcommand{\bbbar}[0]{\ensuremath{b \bar b}\xspace}
\newcommand{\tautau}[0]{\ensuremath{\tau^+ \tau^-}\xspace}
\newcommand{\mumu}[0]{\ensuremath{\mu^{+}\mu^{-}}\xspace}
\newcommand{\ww}[0]{\ensuremath{W^{+}W^{-}}\xspace}

\newcommand{\sigmav}[0]{\ensuremath{\langle \sigma v \rangle}\xspace}

\def\ie{i.e.}

\newcommand{\unit}[1]{\ensuremath{\mathrm{\,#1}}\xspace}

\newcommand{\GeV}{\unit{GeV}}
\newcommand{\cm}{\unit{cm}}
\newcommand{\sr}{\unit{sr}}
\newcommand{\second}{\unit{s}}

\usepackage{lineno}
\begin{document}
\vspace*{4cm}
\title{Constraints on Dark Matter and Supersymmetry from LAT Observations of Dwarf Galaxies}

\author{ Alex Drlica-Wagner on behalf of the Fermi LAT Collaboration }

\address{Kavli Institute for Particle Astrophysics and Cosmology, Department of Physics and SLAC National Accelerator Laboratory, Stanford University, Stanford, CA 94305, USA
}

\maketitle
\abstracts{
Due to a large mass-to-light ratio and low astrophysical backgrounds, dwarf spheroidal galaxies (dSphs) are considered to be one of the most promising targets for dark matter searches via $\gamma$ rays. The $Fermi$ LAT Collaboration has recently reported robust constraints on the dark matter annihilation cross section from a combined analysis of 10 dSphs. These constraints have been applied to experimentally valid, super-symmetric particle models derived from a phenomenological scan of the Minimal Supersymmetric Standard Model (the pMSSM). Additionally, the LAT Collaboration has searched for spatially extended, hard-spectrum $\gamma$-ray sources lacking counterparts in other wavelengths, since they may be associated with dark matter substructures predicted from simulations.
}

\section{Introduction}
It has been well-established that non-baryonic cold dark matter (DM) makes up approximately 85\% of the matter density of the Universe. A popular DM candidate is a weakly interacting massive particle (WIMP) that could pair-annihilate to produce $\gamma$ rays. The $\gamma$-ray flux from WIMP annihilation is given by $\phi(E,\psi)=\sigmav/(8\pi m_W^2)\times N_W(E)\times J(\psi)$, where $\sigmav$ is the velocity averaged pair annihilation cross section, $m_W$ is the WIMP mass, $N_W(E)$ is the $\gamma$-ray energy distribution per annihilation, and  $J(\psi)=\int_{l.o.s.,\Delta\Omega}dl \,d\Omega\rho^{2}[l(\psi)]$ is the line-of-sight (l.o.s.) integral of the squared DM density, $\rho$, toward a direction of observation, $\psi$, integrated over a solid angle, $\Delta\Omega$. Local enhancements in the DM density with large $J(\psi)$, or J-factors, and little astrophysical contamination are potentially good targets for DM searches in $\gamma$ rays. 

The LAT, the primary instrument on board the $Fermi$ observatory, is a pair-conversion telescope with unprecedented sensitivity to $\gamma$ rays in the energy range from $20$ MeV to $>300$ GeV. Scanning the entire sky every three hours, the LAT is an ideal instrument to search for faint new $\gamma$-ray sources. Here we report on the recent LAT searches for Galactic DM substructures that have led to some of the tightest constraints on DM annihilation into $\gamma$ rays.\cite{dwarfs,pmssm,satellites}

\section{Combined Analysis of Dwarf Spheroidal Galaxies\,\protect\cite{dwarfs}}
The Milky Way dSphs are promising sources for the indirect detection of DM via $\gamma$ rays. Stellar velocity data from these galaxies suggest large DM content, while observations at other wavelengths show no signs of astrophysical signals.\cite{Mateo:1998wg,Grcevich:2009gt} With two years of LAT observations, we constrained the $\gamma$-ray signal from ten dSphs using a joint likelihood analysis.\cite{dwarfs}

To set limits on the DM annihilation cross section, we calculated the integrated J-factor within a cone of solid angle $\Delta\Omega = 2.4\times10^{-4} \sr$ centered on each dSph assuming that the DM distribution follows a NFW profile.\cite{Navarro:1996gj} For many of the dSphs, significant statistical uncertainty in the integrated J-factor exists due to limited stellar kinematic data. We incorporated this statistical uncertainty as nuisance parameters in our likelihood formulation:
\begin{equation}
\label{eqn:likelihood}
\begin{aligned}
  L(D\,|\,{\bf p_m},\{{\bf p_k} \}) = \prod_k &L_k^{\rm LAT}(D_k\,|\,{\bf p_m},{\bf p_k})  \\ 
                                              &\times \frac{1}{\ln(10) J_k \sqrt{2 \pi} \sigma_k} e^{-(\log_{10}(J_k)-\overline{\log_{10}(J_k)})^2/2\sigma_k^2} . 
\end{aligned}
\end{equation}
where $k$ indexes the dSphs, $L^{\rm LAT}_k$ denotes the standard LAT binned Poisson likelihood for the analysis of a single dSph, $D_k$ represents the binned $\gamma$-ray data, \{${\bf p_m}$\} represents the dark matter spectral model parameters shared across the dSphs, and \{$\bf{p_k}$\} are dSph-dependent model parameters. 

\begin{figure}
\includegraphics[width=\columnwidth]{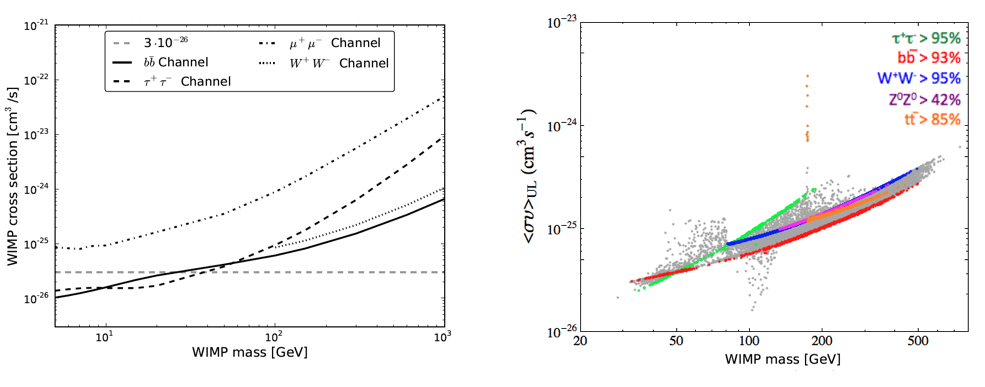}
\caption{Left: Derived 95\% C.L. upper limits on a WIMP annihilation cross section for the \bbbar, \tautau, \mumu, and \ww channels from Ackermann et al. (2011). The generic annihilation cross section ($\sim 3 \times 10^{-26} \,{\rm cm}^3 {\rm s}^{-1}$) is plotted as a reference. Uncertainties in the J factor are included. Right: Similar to the left plot but with each point corresponding to a specific pMSSM model. Colored points denote nearly pure annihilation channels corresponding to the lines in the left plot.
\protect\label{UL_J}}
\end{figure}

In Fig.~\ref{UL_J}, we show combined limits assuming DM annihilation through the \bbbar, \tautau, \ww, and \mumu channels. For the first time, $\gamma$-ray data are able to constrain models possessing the most generic cross section ($\sim 3 \times 10^{-26} \,\cm^3\,\second^{-1}$ for a purely s-wave cross section), without assuming additional boost factors. This strong constraint extends up to a mass of $\sim 27$ GeV for the \bbbar channel and up to a mass of $\sim 37$ GeV for the \tautau channel.

In addition to statistical uncertainties, systematic uncertainties arise from the choice of DM profile. We recalculated our combined limits using the J-factors presented in Charbonnier et al. (2011),\cite{Charbonnier} which allow for shallower profiles than we assumed. We find that the systematic uncertainty resulting from the choice of profile is subdominant, and the combined constraints agree within $\sim10\%$.

\section{Constraints on the pMSSM from Dwarf Spheroidal Galaxies\,\protect\cite{pmssm}}
Supersymmetry (SUSY) is one of the most widely-studied theoretical frameworks for physics beyond the Standard Model (SM). Generic predictions of SUSY are difficult to obtain, and we examined the impact of the dSph limits on a phenomenological subset of the MSSM, the pMSSM.\cite{Berger:2008cq} The pMSSM models obey existing experimental constraints and exhibit a much broader array of phenomenology than can be seen in highly-constrained (mSUGRA/CMSSM) models. The Lightest Supersymmetric Particles (LSPs) of the pMSSM are viable candidates to comprise some or all of DM, and they may be probed through a variety of experimental approaches.

We model the putative $\gamma$-ray emission from the dSphs having spectra generated from $\sim 71$k pMSSM models rather than from prototypical annihilation channels (\ie, \bbbar, \tautau, \mumu, and \ww). We calculated a joint likelihood for each pMSSM model by tying the pMSSM model parameters across the analysis regions surrounding the ten dSphs and incorporating statistical uncertainties in the J-factors of the dSphs. We found no significant $\gamma$-ray signal from any of the dSphs when analyzed individually or jointly for any of the pMSSM models.

\begin{figure}
\includegraphics[width=\columnwidth]{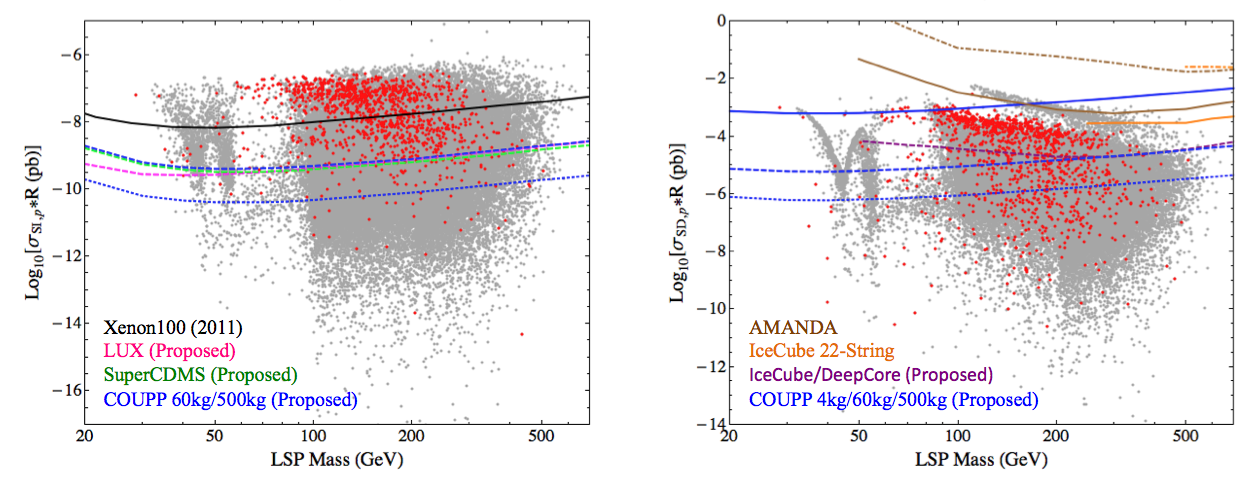}
\caption{Comparison of direct detection limits and LAT dSph constraints from Cotta, et al. (2012). Spin-independent (left) and spin-dependent (right) direct detection cross sections for the pMSSM models are displayed as the gray points, highlighting those within reach of the LAT in red. Limits from current and near-future experiments are displayed as colored lines. Current spin-dependent scattering limits from the AMANDA and IceCube-22 collaborations are displayed with the assumption of soft (dashed) or hard (solid) channel annihilations.
\label{dd}}
\end{figure}

Within the time frame of an extended mission, the LAT has the potential to constrain many of the pMSSM models. In Fig.~\ref{dd}, we display the set of pMSSM models in the spin-independent (left panel) and spin-dependent (right panel) scattering cross section vs.\ LSP mass planes, highlighting the models within reach of future LAT dSph analyses. We observe that many models are expected to be discovered or excluded by both direct detection experiments and the LAT. This could allow for a more detailed characterization of the DM particle. Additionally, we observe that there exist a number of models that will only be accessible to the LAT. These are models whose LSPs are dominantly bino and whose particle spectrum is somewhat hierarchical. These models include a light bino and one or more light sleptons, making them essentially invisible to both direct detection experiments and the LHC due to a lack of accessible colored production channels.

\section{Search for Unassociated Dark Matter Satellites\,\protect\cite{satellites}}

Cosmological N-body simulations predict the existence of many more DM satellites than are observed as dSphs.\cite{Diemand2007,Springel2008} These satellites may be detected as $\gamma$-ray sources having hard spectra, finite angular extents, and no counterparts at other wavelengths.  We selected unassociated, high-Galactic-latitude $\gamma$-ray sources from both the First LAT Source Catalog (1FGL)\,\cite{1FGL} and from an independent list of source candidates created with looser assumptions on the source spectrum.  Using the likelihood ratio test, we distinguished extended sources from point sources and WIMP annihilation spectra from conventional power-law spectra. No candidates were found in either the unassociated 1FGL sources or our additional list of candidate sources.  This null detection is combined with the Via Lactea II~\cite{Diemand2007} and Aquarius~\cite{Springel2008} simulations to set an upper limit on the annihilation cross section for a $100\GeV$ WIMP annihilating through the \bbbar channel.

Using the detection efficiency of our selection, the absence of DM satellite candidates can be combined with the Aquarius and Via Lactea II simulations to constrain a conventional $100\GeV$ WIMP annihilating through the \bbbar channel. We calculated the probability of detecting no satellites from the individual detection efficiency of each simulated satellite. By increasing the satellite flux until the probability of detecting no satellites drops below 5\%, we set a $95\%$ confidence upper limit on \sigmav to be less than $1.95 \times 10^{-24} \cm^{3} \second^{-1}$ for a $100\GeV$ WIMP annihilating through the \bbbar channel.\cite{satellites}

\section{Conclusion}
LAT observations of Galactic DM substructure have placed some of the most robust constraints on the annihilation cross section to date. Observations of known dSphs place tight constraints that can be extended to a broader category of supersymmetric models presented by the pMSSM. Additionally, the null detection of DM substructures as unassociated, spatially-extended, hard-spectrum $\gamma$-ray sources can be combined with predictions from simulations to place independent limit on DM annihilation to \bbbar. In the future, each of these studies will benefit from increased observation time and improvements to the LAT performance.

\section*{Acknowledgments}
The $Fermi$ LAT Collaboration acknowledges support from a number of agencies and institutes for both development and the operation of the LAT as well as scientific data analysis. These include NASA and DOE in the United States, CEA/Irfu and IN2P3/CNRS in France, ASI and INFN in Italy, MEXT, KEK, and JAXA in Japan, and the K.~A.~Wallenberg Foundation, the Swedish Research Council and the National Space Board in Sweden. Additional support from INAF in Italy and CNES in France for science analysis during the operations phase is also gratefully acknowledged.

ADW is supported in part by the Department of Energy Office of Science Graduate Fellowship Program (DOE SCGF), made possible in part by  the American Recovery and Reinvestment Act of 2009, administered by ORISE-ORAU under contract no. DE-AC05-06OR23100.

\end{document}